\documentclass[a4paper,11pt]{article}
\usepackage{pos}

\usepackage{amsfonts}
\usepackage{amsmath}

\newcommand{\Ac}{\mathcal{A}}
\newcommand{\Gc}{\mathcal{G}}
\newcommand{\Jc}{\mathcal{J}}
\newcommand{\Kc}{\mathcal{K}}
\newcommand{\Mc}{\mathcal{M}}
\newcommand{\Oc}{\mathcal{O}}
\newcommand{\Qc}{\mathcal{Q}}
\newcommand{\Rc}{\mathcal{R}}
\newcommand{\Wc}{\mathcal{W}}

\newcommand{\2}{\mathbf{2}}

\DeclareMathOperator{\tr}{Tr}

\newcommand{\df}{\textrm{df}}
\newcommand{\nn}{\nonumber}
\newcommand{\diff}{\textrm{d}}

\title{Connecting Matrix Elements to Multi-Hadron Form-Factors}
\ShortTitle{Connecting Matrix Elements to Multi-Hadron Form-Factors}

\author*[a,b]{Andrew W. Jackura}
        
\affiliation[a]{Thomas Jefferson National Accelerator Facility, 
12000 Jefferson Avenue, Newport News, VA 23606, USA \\}

\affiliation[b]{Department of Physics, 
Old Dominion University, 
Norfolk, Virginia 23529, USA }

\emailAdd{ajackura@odu.edu}

\abstract{We discuss developments in calculating multi-hadron form-factors and transition processes via lattice QCD. Our primary tools are finite-volume scaling relations, which map spectra and matrix elements to the corresponding multi-hadron infinite-volume amplitudes. We focus on two hadron processes probed by an external current, and provide various checks on the finite-volume formalism in the limiting cases of perturbative interactions and systems forming a bound state. By studying model-independent properties of the infinite-volume amplitudes, we are able to rigorously define form-factors of resonances. \\

Preprint number: JLAB-THY-21-3522
}

\FullConference{%
 The 38th International Symposium on Lattice Field Theory, LATTICE2021
  26th-30th July, 2021
  Zoom/Gather@Massachusetts Institute of Technology
}

\begin{document}
\maketitle

\section{Introduction}

Understanding the composition and structure of the excited states of Quantum Chromodynamics (QCD) requires not only constructing model independent observables which can be rigorously attached to an interpretation, but also connecting the excited states via their decays to multi-hadron systems. Lattice QCD, combined with principles of scattering theory and general features of effective field theories (EFTs), offers a systematic pathway to compute excited states from QCD. Since we are interested in capturing the few-body effects of such states, we require a non-perturbative framework to map finite volume observables computed with lattice QCD to the infinite volume scattering amplitudes, from which we can determine the spectrum of the excited states of such reaction processes. 

The L\"uscher methodology provides such a framework, where one exploits the fact that poles of the correlation function, corresponding to the finite volume spectrum, can be related to scattering amplitudes when writing the correlation function to all-orders in the strong coupling~\cite{Luscher:1986pf, Luscher:1990ux,Rummukainen:1995vs, Kim:2005gf, He:2005ey, Hansen:2012tf, Briceno:2014oea}. Applications of the L\"uscher framework to the scattering of two particles have been extremely successful, including cases of particles with spin and multiple coupled channels (see Ref.~\cite{Briceno:2017max} for a review). Generalizations have been studied for transitions from a single hadron to two hadron states~\cite{Lellouch:2000pv,Briceno:2014uqa,Briceno:2015csa,Shultz:2015pfa,Briceno:2021xlc}, including applications to cases such as $\pi+\gamma^{\star}\to \pi\pi$ in the $\rho$-resonance channel~\cite{Briceno:2015dca,Briceno:2016kkp,Alexandrou:2018jbt}.

\begin{figure*}
	\centering
	\includegraphics[width=1.0\textwidth]{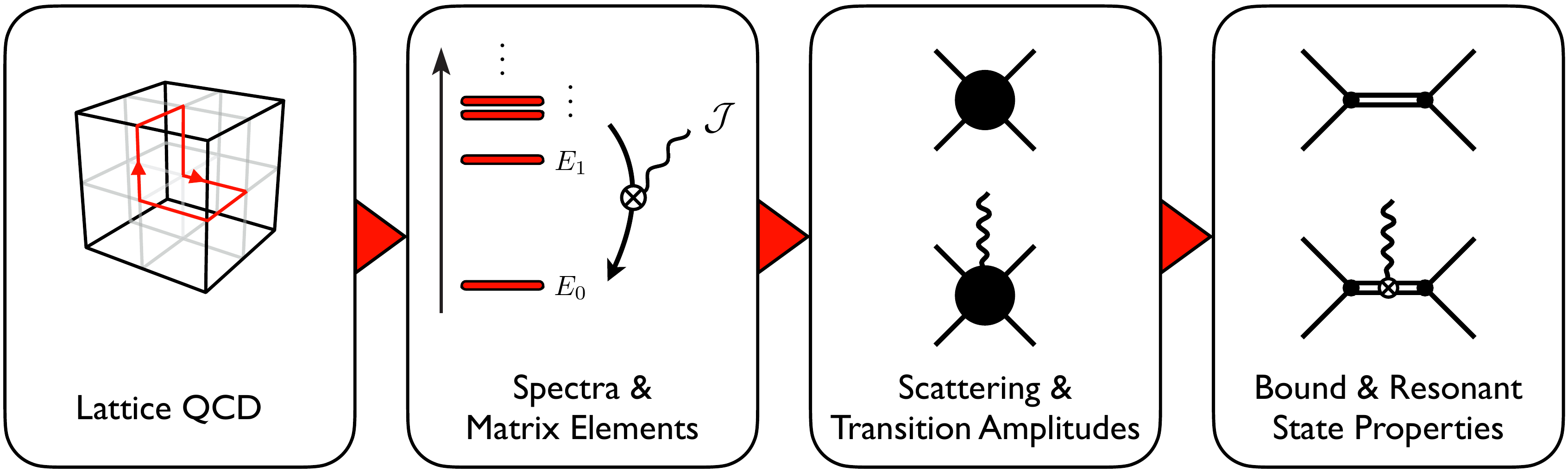}
	\caption{Schematic work flow for connecting finite volume spectra and matrix elements of multiple hadrons computed with lattice QCD to infinite volume bound and resonant state properties.}
	\label{fig:roadmap}
\end{figure*}

Recent extensions~\cite{Agadjanov:2016fbd,Briceno:2012yi,Briceno:2015tza,Baroni:2018iau} have opened up the possibility of studying two-hadron bound and resonant states under the influence of some external current~\cite{Albaladejo:2012te}. The simplest structural observable we investigate are form-factors, which provide the first quantitative information on the size of these excited states, e.g. electromagnetic charge radii of resonances such as $\rho+\gamma^{\star}\to\rho$. Figure~\ref{fig:roadmap} shows the workflow in determining properties of two-body hadronic states. Here we first define our observables of interest, and then review the connection to finite-volume matrix elements and various consistency checks. Then we connect the corresponding infinite-volume observables to an on-shell representation which allow us to perform global analyses using lattice QCD data.

\section{Bound and resonant states of two-hadron systems}

\begin{figure*}[t!]
	\centering
	\includegraphics[width=0.6\textwidth]{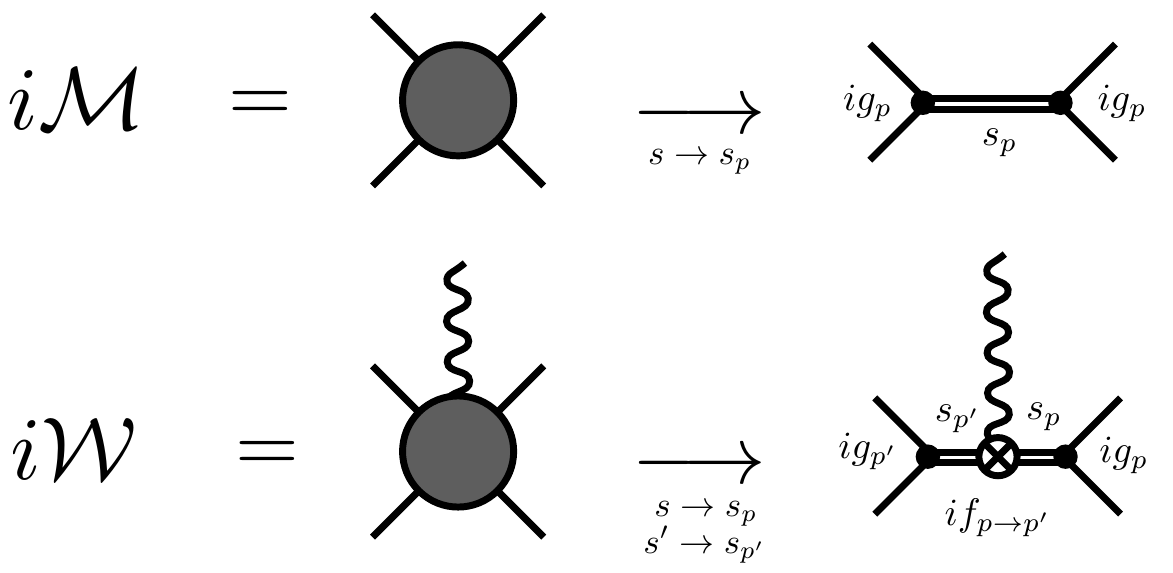}
	\caption{Diagrammatic representations of the $\2\to\2$ and $\2+\Jc\to\2$ amplitudes, defined in Eqs.~\eqref{eq:M_amp} and \eqref{eq:W_amp}, respectively, and representations near bound or resonant state poles as defined in Eqs.~\eqref{eq:bound_resonant} and \eqref{eq:bound_resonant_ff}.}
	\label{fig:amps}
\end{figure*}

We are interested in the spectrum of multi-hadron systems, which is encoded in their  scattering amplitudes. For a $\2\to\2$ scattering system with initial state having four-momentum $P=(E,\mathbf{P})$ and final state having $P'=(E',\mathbf{P}')$, the scattering amplitude $\Mc$ is defined as the asymptotic matrix element
\begin{align}
\label{eq:M_amp}
	\langle P', \mathrm{out}\lvert P , \mathrm{in} \rangle_{\mathrm{conn.}} \equiv (2\pi)^{4}\delta^{(4)}(P'-P) \, i\Mc(P) \, ,
\end{align}
where the delta function enforces total momentum conservation, and we have implicitly assumed states in a definite angular momentum. The ``$\mathrm{conn}.$'' reminds us that only the connected pieces contribute to the amplitude. Since we are interested in the form-factors of scattering states, we also introduce the $\2+\Jc\to\2$ amplitude $\Wc$, where $\Jc$ is some external current with an arbitrary Lorentz structure. We define this amplitude via the matrix element
\begin{align}
\label{eq:W_amp}
	\langle P',\mathrm{out} \rvert \Jc(0) \lvert P, \mathrm{in}\rangle_{\mathrm{conn.}} \equiv \Wc(P',P) \, .
\end{align}
Since the current can inject a momentum $P'-P$, $P\ne P'$ in general and the angular momentum between the initial and final state need not be conserved as well. Although the framework present holds for arbitrary currents, we will focus on the case of an external vector current unless otherwise stated. Figure~\ref{fig:amps} illustrates these amplitudes diagrammatically.

\begin{figure*}[t!]
	\centering
	\includegraphics[width=0.6\textwidth]{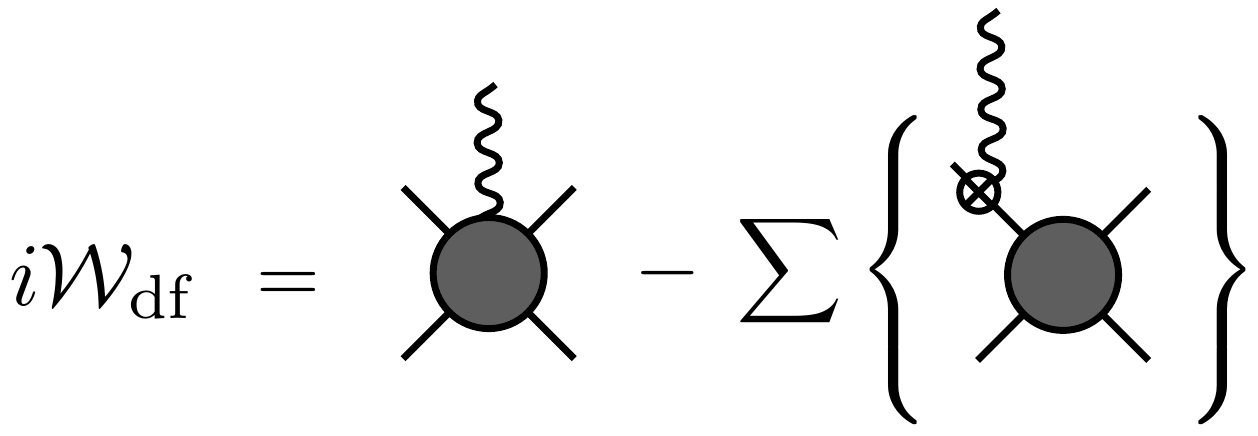}
	\caption{Definition of the divergence free $\Wc_{\df}$, where contributions with current insertions on one of the external legs are removed. For details, we refer to Refs.~\cite{Briceno:2015tza,Baroni:2018iau,Briceno:2020vgp}}
	\label{fig:Wdf}
\end{figure*}

States of multi-hadron systems are characterized by pole singularities of scattering amplitudes. Bound states, such as the deuteron in $NN$, are poles of the amplitude below the scattering threshold, while resonant or virtual states are poles on the analytically continued amplitude to the second Riemann sheet~\footnote{This holds for single channel scattering. For coupled channels, resonances lie on the nearest unphysical sheet to the real energy region under study.}
\begin{align}
\label{eq:bound_resonant}
	\Mc(P) \sim \frac{(ig_p)^2}{s - s_{p}} \quad \mathrm{as\,\,} s\to s_p \, ,
\end{align}
where $s\equiv P^2$,  $s_p$ is the pole position, and $g_p$ is the coupling to the channel. In the same manner, we can define a form factor of a multi-hadron state by continuing the $\Wc$ amplitude to the appropriate Riemann sheet, and isolating the pole contributions, 
\begin{align}
\label{eq:bound_resonant_ff}
	\Wc^{\mu}(P',P) \sim \Wc_{\df}^{\mu}(P',P) \sim \frac{ig_{p'}}{s'-s_{p'}} \, \sum_{j} K_j^{\mu}\, f_{p\to p'}^{(j)}(Q^2) \, \frac{ig_p}{s-s_p} \quad \mathrm{as\,\,} s'\to s_{p'} \,\,\mathrm{and\,\,} s\to s_p\, ,
\end{align}
where $Q^2 = -(P'-P)^2$ is the invariant momentum transfer to the system via the external current, and $\Wc_{\df}$ removes the contributions where the external current couples to one of the external legs infinite volume $\2+\Jc\to \2$ amplitude $\Wc$, as discussed in Refs.~\cite{Briceno:2015tza,Baroni:2018iau,Briceno:2020vgp} and shown in Fig.~\ref{fig:Wdf}, Note that the terms which contain the current insertion on one of the external legs do not contribute as one of the two spectral poles will cancel against this term, meaning the spectral poles can be determined from $\Wc_{\df}$ alone. Here, $p$ and $p'$ are the states in the spectrum, $K_j$ is a kinematic function determining the Lorentz structure, and $f_{p\to p'}^{(j)}$ are scalar form factors. Elastic form factors are those where the state is identical, $p\to p$, and transitions are for $p\ne p'$. This allows us to provide rigorous definitions for elastic resonance form factors $f_{R\to R}$, for example $\rho+\gamma^{\star}\to\rho$ which in turn can give us the electromagnetic charge radius and provide first insight to a structural observable of a resonance from QCD. See Fig.~\ref{fig:amps} for a pictorial representation of Eqs.~\eqref{eq:bound_resonant} and \eqref{eq:bound_resonant_ff}.

\section{Two-hadron matrix element finite volume formalism}

We consider a cubic volume $L^3$ subject to periodic boundary conditions with an infinite temporal extent $T\to\infty$. The primary objects we consider are the 2- and 3-point correlation functions, which can be related to the finite volume spectrum and matrix elements. For some interpolating operator $\Oc$, the 2-point correlator is given by
\begin{align}
	C_{2,L}(P) = \int_L\! \diff^4 x \, e^{iP\cdot x} \langle \Oc(x)\Oc^{\dag}(0) \rangle = L^3 \sum_{\mathfrak{n}}\frac{i Z_{\mathfrak{n}}Z_{\mathfrak{n}}^{\dag}}{E - E_{\mathfrak{n}} + i\epsilon} \, ,
\end{align}
where $x = (t,\mathbf{x})$ is a Minkowski spacetime point and we assume a time ordering $t>0$ and $Z_{\mathfrak{n}} \equiv \langle 0 \rvert \Oc(0) \rvert P_{\mathfrak{n}},L \rangle$ is an overlap factor for the interpolating operator onto some finite volume state $\mathfrak{n}$ with $P_{\mathfrak{n}} = (E_{\mathfrak{n}},\mathbf{P})$. The second equality is the spectral decomposition of the correlator, which shows that the spectrum $E_{\mathfrak{n}}$ for the system arises as pole singularities on the real energy axis of the function. In practice, one computes these correlation functions in Euclidean time $t \to -i\tau$. Here, we simply work with Minkowski time $t$ since the spectrum and matrix elements considered here are independent of the time signature. Similarly, the $3$-point function is defined by
\begin{align}
	C_{3,L}(P',P) & = \int_L\! \diff^4 x' \int_L\!\diff^4 x \, e^{iP'\cdot x'}e^{-i(P'-P)\cdot x} \langle \Oc(x')\Jc(x)\Oc^{\dag}(0) \rangle \, , \nn \\[5pt]
	& = L^6 \sum_{\mathfrak{m},\mathfrak{n}} \frac{iZ_{\mathfrak{m}}}{E'-E_{\mathfrak{m}}+i\epsilon} \, \langle P_{\mathfrak{m}}', L\rvert \Jc(0)\lvert P_{\mathfrak{n}},L\rangle \, \frac{iZ_{\mathfrak{n}}^{\dag}}{E-E_{\mathfrak{n}}+i\epsilon} \, ,
\end{align}
where it is assumed that $t' > t > 0$ and $\langle P_{\mathfrak{m}}', L\rvert \Jc(0)\lvert P_{\mathfrak{n}},L\rangle$ is the matrix element between two finite volume states $\mathfrak{n}$ and $\mathfrak{m}$. For some particular set of interpolating operators with the desired quantum numbers, one computes these correlation functions and extracts the finite volume spectrum and matrix elements.

\begin{figure*}
	\centering
	\includegraphics[width=0.98\textwidth]{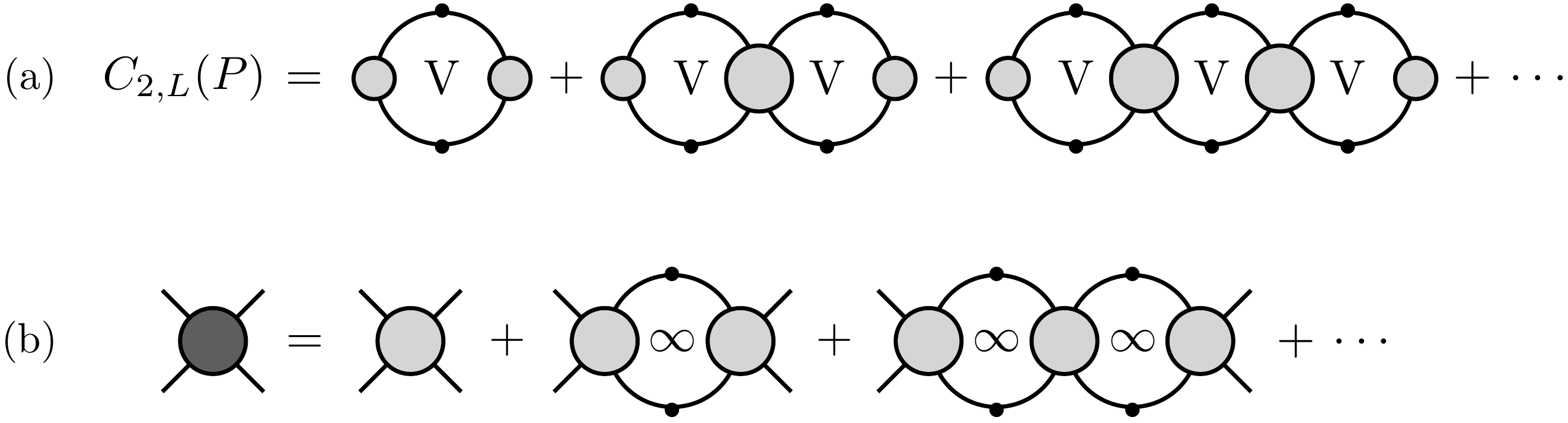}
	\caption{(a) Diagrammatic representation for the 2-point correlator in terms of the fully dressed propagator (line with black dot), the Bethe-Salpeter two-particle irreducible kernel (gray circle with four legs), and the end-cap kernels. Each loop labeled with `$\mathrm{V}$' is an integral over the temporal component of momentum and a sum over the spatial. (b) The infinite-volume $\2\to\2$ amplitude in terms of the Bethe-Salpeter kernel.}
	\label{fig:all_orders}
\end{figure*}

In addition to the spectral representations, we can write expressions for the 2- and 3-point correlation functions to all-orders within some generalized EFT. Since we are interested in the infrared limit of QCD, that is hadronic processes, the generalized EFT provides an intermediary tool to encapsulate the hadronic degrees of freedom. We do not specify any particular interaction, and allow any and all interactions which satisfy the symmetries of the system under consideration. Within this framework, one can write diagrammatic representations of the correlators in terms of the fully dressed propagators of the hadrons, and one- and two-particle irreducible kernels that encode all the physics which does not allow two-particle states to go on-shell. Figure~\ref{fig:all_orders} illustrates an example of the 2-point correlator for two-particle systems. 

Identifying the poles of the all-orders expression with the spectral representation allows us to connect finite-volume and infinite-volume quantities. For example, if we consider a system such that the ground state of some function is associated with the finite volume mass of the hadron, $M_L = E_{\mathfrak{0}}$, then when we express all finite-volume sums in the loops as integrals plus a correction determined from the Poisson summation formula and find that the finite volume mass of a hadron is equal to its infinite volume mass $M$ up to some exponentially small correction,
\begin{align}
	M_L = M + \Oc(e^{-mL}) \, ,
\end{align}
where $m$ is the smallest mass scale of the theory, e.g. the $\pi$ for QCD. We assume we work in a regime such that $mL \gg 1$, thus this correction is negligible compared to other volume effects which we consider shortly. Besides matching pole positions, we may also match the residues at the poles between the spectral and all-orders representations. For the 3-point function, this allows us to link matrix elements with form factors, e.g. for the elastic form factor of a stable hadron, one finds that the form factors $f_L$ are equal to their infinite-volume counterparts $f$ up to some exponentially suppressed correction
\begin{align}
	f_L(Q^2) = f(Q^2) + \Oc(e^{-mL}) \, ,
\end{align}
where $Q^2 = -(P'-P)^2$ is the momentum transfer.

Finite volume corrections for multi-hadron systems scale as a power law of the inverse volume, since on-shell intermediate states lead to singularities in the correlation function which enhance the volume dependence. Within a kinematic region where only two particles can go on their mass-shell, i.e. $2m \le E^{\star} < E_{\mathrm{inel.}}^{\star}$ where $E^{\star} \equiv \sqrt{P^2}$ is the center-of-momentum (CM) energy of the system and $E_{\mathrm{inel.}}^{\star}$ is the first inelastic threshold, the finite volume correction to the correlator leads to the L\"uscher quantization condition, which relates the poles of the correlator to the scattering amplitude $\Mc$,
\begin{align}
\label{eq:luscher}
	\det\left[ \, \mathcal{M}^{-1}(P_{\mathfrak{n}}) + F_L(P_{\mathfrak{n}}) \, \right] = 0 \, ,
\end{align}
where the determinant is over angular momentum space and $P_\mathfrak{n} = (E_{\mathfrak{n}},\mathbf{P})$ is the four-momentum of the two-particle system in a frame with momentum $\mathbf{P}$ and finite volume energy $E_{\mathfrak{n}}$. Here, $F_L$ is a known function characterizing the geometry of the volume, and is a dense matrix in angular momentum space. For example, considering identical particles in $S$ wave, $F_L$ takes the form
\begin{align}
	F_L(P) = \xi \left[\, \frac{1}{L^3} \sum_{\mathbf{k}} - \int\! \frac{\diff^3 \mathbf{k}}{(2\pi)^3} \, \right] \, \frac{1}{2\omega_k} \frac{1}{(P-k)^2 - m^2+i\epsilon} \bigg\rvert_{k^0 = \omega_{k}} \, ,
\end{align}
where $\xi = 1/2$ for identical particles and the sum over $\mathbf{k}$ is over all $\mathbf{k} = 2\pi \mathbf{n} / L$ where $\mathbf{n} \in \mathbb{Z}^3$. Equation~\eqref{eq:luscher} can be subduced into irreducible representation of the Octahedral group or its little groups for systems at rest or moving, respectively.

\begin{figure*}
	\centering
	\includegraphics[width=0.98\textwidth]{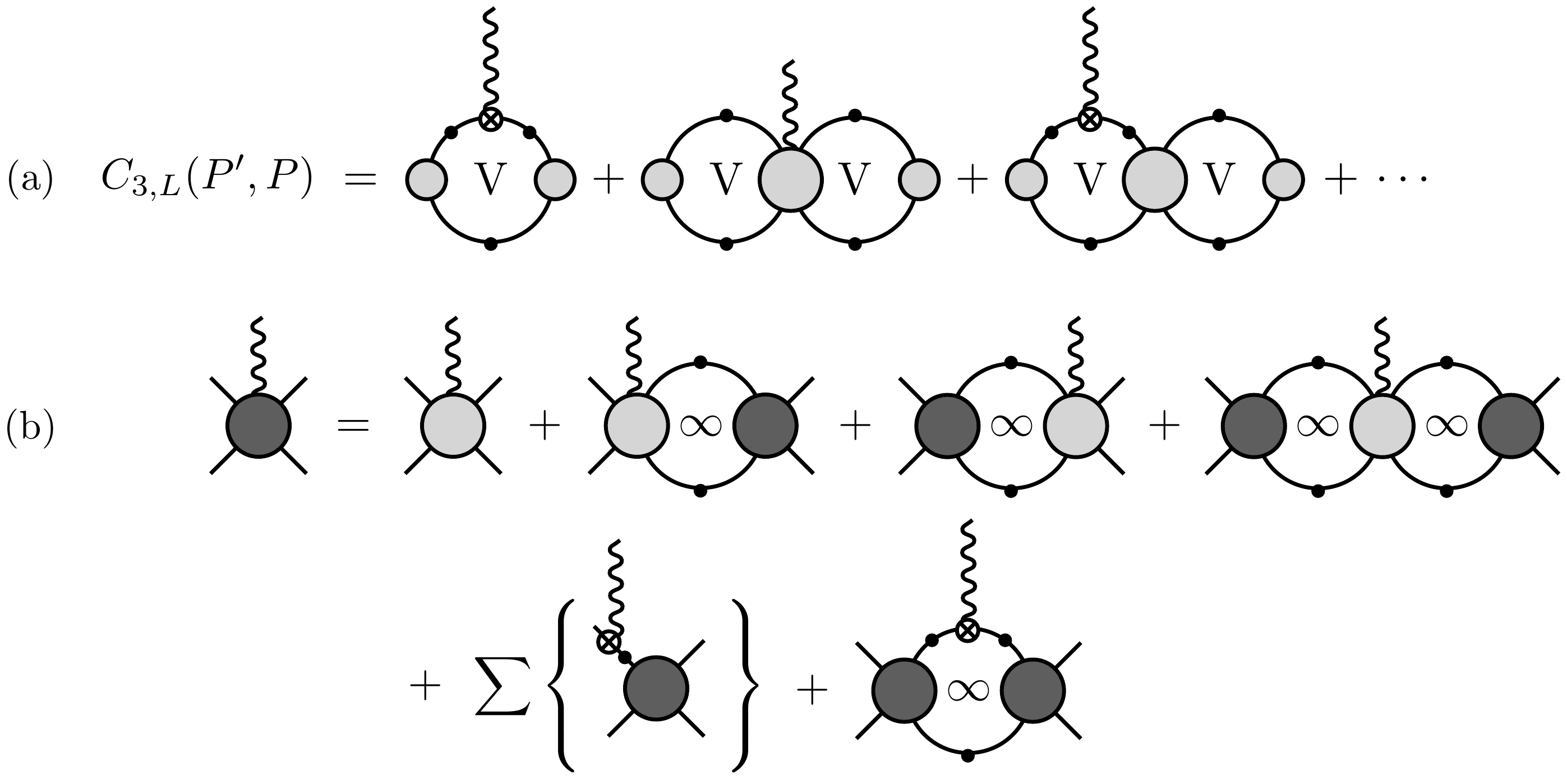}
	\caption{(a) Diagrammatic expansion for the 3-point correlator, with short-distance kernels for the single current insertions on single particles (cross) and on two-particle irreducible kernels (gray circle with four legs). (b) The all-orders representation for the infinite-volume $\2+\Jc\to\2$ amplitude in terms of its kernels and the $\2\to\2$ amplitude.}
	\label{fig:all_orders_2}
\end{figure*}

Turning to $\2+\Jc\to\2$ processes, cf. Fig.~\ref{fig:all_orders_2}, examining simultaneous residues of the finite volume correction, one can identify a relation between the matrix element and the corresponding infinite volume amplitude~\cite{Briceno:2015tza,Baroni:2018iau}
\begin{align}
\label{eq:briceno_hansen}
	\left\lvert \, \langle P_{\mathfrak{m}}', L\rvert \Jc^{\mu}(0)\lvert P_{\mathfrak{n}},L\rangle \, \right\rvert^2 = \frac{1}{L^6}\tr\left[ \, \Wc_{L,\df}^{\mu}(P_{\mathfrak{n}},P_{\mathfrak{m}}') \cdot \Rc_L(P_{\mathfrak{m}}') \cdot \Wc_{L,\df}^{\mu}(P_{\mathfrak{m}}',P_{\mathfrak{n}}) \cdot \Rc_L(P_{\mathfrak{n}}) \, \right]
\end{align}
where $\Rc_L$ is the generalized Lellouch-L\"uscher factor~\cite{Briceno:2014uqa,Briceno:2015csa},
\begin{align}
	\Rc_L(P_{\mathfrak{n}}) = \lim_{E\to E_{\mathfrak{n}}} \, \frac{(E - E_{\mathfrak{n}}) }{\Mc^{-1}(P) + F_L(P)} \, ,
\end{align}
and $\Wc_{L,\df}$ is defined as
\begin{align}
	\Wc_{L,\df}^{\mu}(P',P) = \Wc_{\df}^{\mu}(P',P) + \Mc(P') \cdot \left[ \, \sum_{j} f_j(Q^2) \, \Gc_{L,j}^{\mu}(P',P) \, \right] \cdot \Mc(P) \, ,
\end{align}
with $f_j$ are the single hadron form factors of the constituent system, and $\Gc_L$ is a geometric function which characterizes the finite-volume corrections for this object. Consider a system with a vector current, with total $S$ wave, then there is a single form factor $f$ and function $\Gc_L$, which takes the form
\begin{align}
	\Gc_L^{\mu}(P',P) = \left[\, \frac{1}{L^3} \sum_{\mathbf{k}} - \int\! \frac{\diff^3 \mathbf{k}}{(2\pi)^3} \, \right] \, \frac{1}{2\omega_k} \frac{(P'+P)^{\mu}-2k^{\mu}}{\left[(P'-k)^2 - m^2+i\epsilon \right]\left[(P-k)^2 - m^2+i\epsilon\right]}  \bigg\rvert_{k^0 = \omega_{k}} \, .
\end{align}
A diagrammatic representation of the infinite-volume portion is shown in Fig.~\ref{fig:triangle}.

\subsection{Volume dependence of conserved vector charge}

As part of a series of consistency checks, we examine some special cases of the finite-volume formalism for the $\2+\Jc\to\2$ processes. First, we consider a conserved vector current $\partial_{\mu} \Jc^{\mu} = 0$. Current conservation places restrictions in the form of the Ward-Takahashi identity, leading to a constraint on the forward direction of the $\Wc_{\df}$ amplitude. For $S$ wave scattering of a system composed of a neutral and charged particle, e.g. $\pi^0\pi^+$, the relation is given by
\begin{align}
	\Wc_{\df}^{\mu}(P,P) = \mathrm{Q} \, \frac{\partial}{\partial P_{\mu}} \Mc(P) \, ,
\end{align}
where $\mathrm{Q} = f(0)$ is the charge of the single particle. This relation ensures that the charge of the matrix element is protected from finite-volume effects~\cite{Briceno:2019nns},
\begin{align}
	\langle P_{\mathfrak{n}}, L \rvert \Qc \lvert P_{\mathfrak{n}}, L \rangle \equiv L^3\langle P_{\mathfrak{n}}, L \rvert \Jc^{0}(0) \lvert P_{\mathfrak{n}}, L \rangle = \mathrm{Q} \, .
\end{align}

\subsection{Volume dependence for matrix elements of bound states}

If we consider energies below the two-body production threshold, then the power-law finite-volume corrections become exponentially suppressed via the Poisson summation formula. If there is a bound state in this vicinity, then the bound state mass will gain exponentially suppressed corrections~\cite{Davoudi:2011md}
\begin{align}
	M_{B,L} = M_B + \Oc\left( e^{-\kappa L} \right) \, ,
\end{align}
where $\kappa$ is the binding momentum. If we examine Eq.~\eqref{eq:briceno_hansen} in this region, we find that a bound state form-factor will gain similar exponentially suppressed corrections. Consider a scalar current, $\Jc_S$, we investigate the volume dependence of the scalar charge and find~\cite{Briceno:2019nns}
\begin{align}
	\mathrm{g}_{B,L} & \equiv 2E_{B,L} \, \langle P_B, L \rvert \Jc_{S}(0) \lvert P_B, L \rangle \, , \nn \\[5pt]
	& = \mathrm{g}_{B} + \Oc\left( e^{-\kappa L} \right) \, ,
\end{align}
where $\mathrm{g}_{B}$ is the infinite-volume scalar charge.

\subsection{Threshold expansion of matrix elements}

As a final check of the formalism, we explore the near threshold behavior of matrix elements as a series in the inverse volume. For identical scalars, the ground state CM energy follows the well-known expansion~\cite{Luscher:1986pf}
\begin{align}
\label{eq:E0_Thresh}
	E_0(L) = 2m + \frac{4\pi a}{mL^3} + \Oc\left(\frac{1}{L^4}\right)  \, ,
\end{align}
where $a$ is the $S$ wave scattering length. One arrives at Eq.~\eqref{eq:E0_Thresh} by looking at the non-relativistic limit of Eq.~\eqref{eq:luscher}, and expand to the desired order. If we now consider a scalar current in the forward direction, we find that following a similar expansion of Eq.~\eqref{eq:briceno_hansen} leads to the expression~\cite{Briceno:2020xxs}
\begin{align}
	L^3 \langle \mathfrak{n}=0, L \rvert \Jc_{S}(0) \lvert \mathfrak{n}=0, L \rangle = \frac{\mathrm{g}}{m}\left[ 1 - \frac{2\pi a}{m^2 L^3} + \Oc\left(\frac{1}{L^4}\right) \right] \, ,
\end{align}
where $\mathrm{g}$ is the scalar charge of a single particle. Alternatively, one can arrive at this expansion by using time-ordered perturbation theory on the 3-point correlation function. This expansion agrees with the Feynman-Hellman theorem, 
\begin{align}
	L^3 \langle \mathfrak{n}=0, L \rvert \Jc_{S}(0) \lvert \mathfrak{n}=0, L \rangle = \mathrm{g} \, \frac{\diff E_0(L)}{\diff m^2} \, ,
\end{align}
where we expect the scaling of the correction to the energy matches that of the matrix element.

\section{Resonant form-factors from on-shell amplitudes}

Given a set of lattice QCD spectra and matrix elements for some given channel, we use Eq.~\eqref{eq:luscher}, along with \eqref{eq:briceno_hansen} to constrain the infinite volume on-shell $\Mc$ and $\Wc_{\df}$. Within a limited region of the CM energy, below inelastic thresholds, the infinite volume amplitudes admit an on-shell representation which isolates the non-analyticities fixed from unitarity, i.e. branch point singularities associated with on-shell intermediate states are separated from short-distance physics which is unknown. The well-known $K$ matrix form for the $\2\to\2$ partial wave amplitude is one such representation, 
\begin{align}
\label{eq:M_onshell}
	\Mc(P) = \Kc(P) \cdot \frac{1}{1 - i\rho(P) \cdot \Kc(P)}
\end{align}
where $\rho$ is the two-body phase space factor characterizing the kinematics of on-shell two-particle states
\begin{align}
	\rho(P) = \frac{\xi q^{\star}}{8\pi E^{\star}} \, ,
\end{align}
with $q^{\star} = \sqrt{P^2/4 - m^2}$. The function $\Kc$ captures all short-distance (off-shell) physics, and is a real function in the kinematic region of consideration. Unitarity alone does not fix the form of $\Kc$, however using Eq.~\eqref{eq:luscher} with some spectrum determined from lattice QCD allows us to determine $\Kc$ in a global analysis with some suitable parameterization respecting the on-shell structure. 

\begin{figure*}
	\centering
	\includegraphics[width=0.5\textwidth]{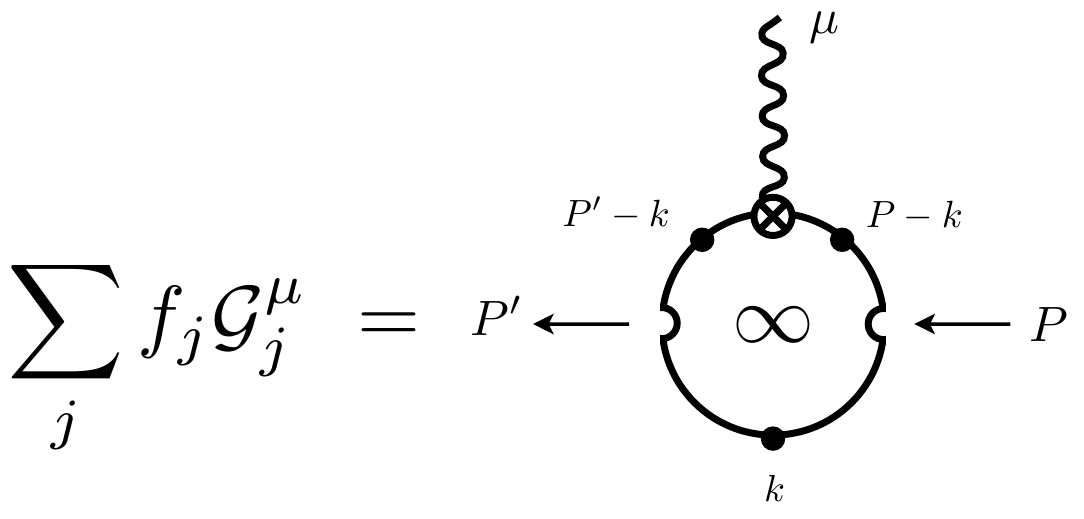}
	\caption{Triangle topology $\sum_j f_j \Gc_j^{\mu}$ for the $\Wc_{\df}$ amplitude shown in Eq.~\eqref{eq:W_onshell}.}
	\label{fig:triangle}
\end{figure*}

In a similar manner to $\Mc$, the $\Wc$ amplitude can be written in an on-shell form, allowing us to parameterize a smooth function and ensure the analytic structure of the function is respected. When we project the all-orders self-consistent equation [Fig.~\ref{fig:all_orders_2}] to an on-shell form, we find that $\mathcal{W}_{\mathrm{df}}$, cf. Fig.~\ref{fig:Wdf}, takes the form~\cite{Briceno:2020vgp}
\begin{align}
\label{eq:W_onshell}
	\mathcal{W}_{\mathrm{df}}^{\mu}(P',P) = \mathcal{M}(P') \cdot \left[\, \mathcal{A}^{\mu}(P',P) + \sum_j f_j(Q^2) \, \mathcal{G}_j^{\mu}(P',P) \, \right] \cdot \mathcal{M}(P) \, ,
\end{align}
where $f_j$ are the on-shell single particle form factors, $\Gc$ is a known kinematic function shown in Fig.~\ref{fig:triangle}, the infinite volume triangle function characterizing on-shell singularities, and $\Ac$ is a smooth real function which encapsulates the short-distance physics of two-particle scattering under the influence of an external current. The $\2\to\2$ amplitudes in the initial and final states describe the initial and final state interactions between the two particles. Again, we have presented the form of Eq.~\eqref{eq:W_onshell} with a vector Lorentz index, but the on-shell form is identical for any Lorentz structure.

The path to bound or resonant state form factors is as follows: First, one computes the two-particle finite volume spectrum for some channel and determines the $K$ matrix and resulting $\2\to\2$ amplitude for a given channel using Eq.~\eqref{eq:luscher} and \eqref{eq:M_onshell}. The single particle form factors $f_j$ of the hadrons in the channel must also be determined for the desired current. Next, the matrix elements for the two hadron system subject to the external current must be calculated, and the $\Wc_{\df}$ amplitude can be constrained by fitting parameterizations of $\Ac$ in Eq.~\eqref{eq:W_onshell} such that \eqref{eq:briceno_hansen} is satisfied. Once an analytic parameterization for $\Ac$ can be determined, the elastic or transition form factors for a bound or resonant state can be determined by inverting Eq.~\eqref{eq:bound_resonant_ff} for $f_{p\to p'}$. In terms of the on-shell representation for $\Wc_{\df}$, the form factors of a state in the two-particle spectrum is given by
\begin{align}
	\sum_{j} K_j^{\mu}\, f_{p\to p'}^{(j)}(Q^2) = \lim_{\substack{s'\to s_{p'} \\ s \to s_p}} g_{p'} g_p \, \left[ \, \Ac^{\mu}(P',P) + \sum_{j} f_j(Q^2) \, \Gc_j^{\mu}(P',P) \, \right] \, ,
\end{align}
where for resonances the expression is defined with $\Gc$ on the second Riemann sheet as detailed in Ref.~\cite{Briceno:2020vgp}. At zero-momentum transfer, the charge of the resonance must be the sum of the constituent hadrons. For a charge-neutral system, e.g. $\rho^+\to \pi^+\pi^0$, we recover $f_{p\to p}(0) = \mathrm{Q}$.

\section{Summary}

Scattering calculations from lattice QCD have matured over the past decade, and the push for developing the framework to compute more complicated observables has evolved in tandem. Including the effects of external current sources interacting with two-body systems has been developed both for the finite volume formalism and the infinite volume amplitude. Non-trivial checks on the formalisms have been made in various cases. Coupled with lattice QCD data, the finite- and infinite-volume formalisms provide a pathway to compute resonance form-factors, providing the first structural observables to study the composition of excited QCD states.

\section{Acknowledgements}
I would like to thank my collaborators Ra\'ul A. Brice\~no, Maxwell T. Hansen, Felipe G. Ortega-Gama, and Keegan H. Sherman for their contributions in completing this work and comments on these proceedings.
This work is supported in part by USDOE grant No. DE-AC05-06OR23177, 
under which Jefferson Science Associates, LLC, manages and operates Jefferson Lab.
This work is also supported from the USDOE Early Career award, contract de-sc0019229.

\end{document}